\documentclass[preprint,showpacs,pra,aps,superscriptaddress,amssymb]{revtex4-1}
\usepackage{amsmath}
\usepackage{amsfonts}
\usepackage{mathrsfs}
\usepackage{amssymb}
\usepackage{bm}
\usepackage{graphicx}
\usepackage{color}
\usepackage[dvipsnames]{xcolor}
\usepackage[colorlinks=true,linkcolor=blue,citecolor=blue,urlcolor=blue]{hyperref}
\begin{document}

\title{Effect of the resonance spectra in the propagation of two decaying entangled particles}
%

%\author{Gast\'on Garc\'{\i}a-Calder\'on$^1$, Roberto Romo$^2$ and  Miguel \'Angel Ter\'an$^1$}
%\address{$^1$Instituto de F\'{\i}sica, Universidad Nacional Aut\'onoma de M\'exico, 04510 Ciudad de M\'exico, Mexico}
%\address{$^2$Facultad de Ciencias, Universidad Aut\'onoma de Baja California, 22800 Ensenada,
%Baja California, Mexico}
%\ead{\mailto{gaston@fisica.unam.mx}}
%
\author{Gast\'on Garc\'{\i}a-Calder\'on}
\email{gaston@fisica.unam.mx}
\affiliation{Instituto de F\'{\i}sica,
Universidad Nacional Aut\'onoma de M\'exico, 04510 Ciudad de M\'exico, M\'exico}
\author{Roberto Romo}
\email{romo@uabc.edu.mx} \affiliation{Facultad de Ciencias, Universidad
Aut\'onoma de Baja California, 22800 Ensenada, Baja California, M\'exico}
\author{Miguel \'Angel Ter\'an}
%\email{gaston@fisica.unam.mx}
\affiliation{Instituto de F\'{\i}sica,
Universidad Nacional Aut\'onoma de M\'exico, 04510 Ciudad de M\'exico, M\'exico}

\begin{abstract}
An exact analytical solution of the decaying wave function of two identical noninteracting particles, which are entangled by spatial symmetry, is used to analyze the effect of the resonance spectra in the propagation of the decaying probability density outside the interaction potential region. We find, using exactly solvable problems, that a usual approximation that considers the two resonance levels associated with the initial states, is affected substantially in the case of sharp high energy resonances by disrupting the pure exponential decaying regime exhibited by the two resonance level approximation, whereas for broad high energy resonances, we find that the probability density profile is well described by the two resonance approximation.
\end{abstract}

%\pacs{03.65.Ca,73.40.
%

\maketitle

\section{Introduction}

In recent work it is shown rigorously, using the analytical properties of the outgoing Green's function for single particle potentials that vanish exactly beyond a distance, that the non-Hermitian formulation using resonance states and the Hermitian formulation in terms of continuum wave solutions to the time dependent Schr\"odinger equation for tunneling decay, yield identical numerical results \cite{gcmv12}. See also  \cite{gcmv07,gcmv13}. The above refers to the dynamics of a class of open quantum systems that deal with the full Hamiltonian to the system in which a particle initially confined within an interaction region decays by tunneling to the outside. This should not be confused with approaches where the Hamiltonian is separated into a part corresponding to a closed system and a part responsible for the decay which usually is treated to some order of perturbation theory, as in the old work by Weisskopf and Wigner to describe the decay  of an excited atom interacting with a quantized radiation field \cite{wigner30}, that has become a standard procedure in some decay problems where perturbation theory may be justified.

Artificial quantum systems at the nanometric scale provide interesting examples of decay of open quantum systems, as electronic decay in double-barrier resonant tunneling semiconductor structures at low temperatures \cite{sakaki87} and the decay of  ultracold atoms initially confined by light \cite{serwane11}. The formalism of resonance states has been used to investigate the conditions for nonexponential decay in these systems \cite{gcv06,gcr16}.

In particular, studies involving ultracold atoms have opened the way to study the decay
of an arbitrary number of atoms. In particular, for two atoms, one may control the interaction strength between the atoms to reach the limit of no interaction between them. These may allow to study the decay of identical non interacting atoms entangled by spatial symmetry. In recent work, Garc\'{\i}a-Calder\'on and Mendoza-Luna showed,
using the formalism of resonant states, that the survival and nonescape probabilities for non interacting identical entangled symmetric and entangled antisymmetric initial states evolve in a distinct form along the exponentially decaying and nonexponential regimes \cite{gcm11}. These probabilities, however, are defined within the potential interaction region both for single and two particle decay, as reviewed in \cite{gc15}, and hence omit the relevant case of propagation of decaying particles beyond the range of the potential.

In this work we investigate the probability density for decay of two identical entangled particles by spatial symmetry as they propagate along the external region of an interaction potential. In particular we address the issue of the effect of the resonance spectra in the time evolution of the decaying probability density.

The paper is organized as follows.  Section 2 reviews briefly the time evolution of decay of a single particle.
Section 3 deals with the decay of symmetric and antisymmetric identical particles and shows that it may be expressed in terms of combinations of single particle decaying density probabilities. Section 4 discusses some model calculations, and finally, section 5  presents some concluding remarks.

\section{Decay of a single particle}

In this section we briefly recall the relevant points of the derivation of the decaying wave solution for a single particle
confined initially, at $t=0$, along the internal region of a real spherically symmetrical potential that vanishes beyond a distance, \textit{i.e.}, $V(r)=0$ for $r>a$. We choose the natural units $\hbar=2m=1$ and for simplicity of the description we refer to $s$ waves.
Hence, the solution to the time-dependent Schr\"odinger equation, as an initial value problem, may be written at time $t > 0$ in terms of the retarded Green's function $g(r,r';t)$ of the problem as \cite{gc10},
\begin{equation}
\Psi(r,t)=\int_0^a {\! g(r,r^\prime,t)\Psi(r',0)\,\mathrm{d}r^\prime},
\label{1s}
\end{equation}
where $\Psi(r,0)$ stands for the arbitrary state initially confined within the internal interaction region.
Since the decay  refers to tunneling into the continuum, for the sake of simplicity it is assumed that the  potential does not possess bound nor antibound states. It is convenient to express the
retarded time-dependent Green's function in terms of the outgoing Green's function $G^+(r,r';k)$ of the problem.
Both quantities are related by a Laplace transformation. The Bromwich contour in the $k$ complex plane  corresponds to a hyperbolic contour along the first quadrant that may be deformed to a contour that goes from $-\infty$ to $\infty$ along the real $k$ axis,
\begin{equation}
g(r,r';t)={i \over 2 \pi} \int_{-\infty}^{\infty} G^+(r,r\,';k) {\rm e}^{-i k^2t} \,2kdk.
\label{74}
\end{equation}
This allows to make use of the resonant expansion of the outgoing Green's function \cite{gc10}
\begin{equation}
G^+(r,r\,';k) = \frac{1}{2k}\sum_{n=-\infty}^{\infty} \frac {u_n(r)u_n(r\,')}{k-\kappa_n}, \quad  (r,r')^{\dagger} \leq a
\label{9x}
\end{equation}
where the notation $(r,r')^\dagger$  means that the point $r=r^\prime=a$ is excluded in the above expansion (otherwise it diverges) and the set of functions $\{u_n(r)\}$ refer to the resonance  states to the problem, which are obtained from the residues at the complex poles $\{\kappa_n\}$ of the outgoing Green's function that also provide its normalization condition \cite{gcp76,gc10},
\begin{equation}
\int_0^a u_n^2(r) dr + i\frac{u_n^2(a)}{2\kappa_n}=1.
\label{74c}
\end{equation}
Resonance states satisfy the Schr\"{o}dinger equation of the problem $[\kappa_n^2-H]u_n(r)=0$ with outgoing (radiative) boundary conditions $u_n(0)=0$,\quad $[du_n(r)/dr]_{r=a}=i\kappa_nu_n(a)$. The complex energy eigenvalues are $\kappa_n^2= E_n=\mathcal{E}_n-i\Gamma_n/2$, where $\mathcal{E}_n$ stands for the resonance energy of the decaying fragment and $\Gamma_n$ stands for the resonance width,  which is related to the lifetime as, recalling that $\hbar=1$, $\tau_n=1/\Gamma_ n$. The lifetime of the system is defined by the longest lifetime. The complex poles $\kappa_n=\upsilon_n-i\gamma_n$ are distributed along the third and fourth quadrants of the complex $k$ plane in a well known fashion \cite{newton}.

One should note that for $r' <a $ and $r \geq  a$ one may write $G^+(r,r\,';k)$ as \cite{gcmv12,gc10},
\begin{equation}
G^+(r,r\,';k)=G^+(r',a;k) {\rm e}^{ik(r-a)}, \quad r'<a, \quad r \geq a.
\label{75c}
\end{equation}
The above expression permits to get a resonance expansion for $G^+(r,r\,';k)$  which is valid along the external region of the interaction. On the other hand, it is easily shown that the representation of $G^+(r,r\,';k)$ given by (\ref{9x}) satisfies the closure relation \cite{gcmv12,gc10},
\begin{equation}
{\rm Re} \left\{\sum_{n=1}^{\infty} u_n(r)u_n(r\,')\right\}=\delta(r-r\,'),\quad  (r,r')^{\dagger} \leq a.
\label{9y}
\end{equation}

The above results allows us to write the decaying time-dependent wave function as \cite{gcmv12,gc10},
\begin{equation}
\Psi(r,t)=\sum_{n=-\infty}^{\infty}
\left \{ \begin{array}{cc}
C_nu_n(r)M(y^\circ_n), & \quad  r \leq a \\[.4cm]
C_nu_n(a)M(y_n), & \quad r \geq a,
\end{array}
\label{b6}
\right.
\end{equation}
where the coefficients $C_n$ are  defined as,
\begin{equation}
C_n=\int_0^a \Psi(r,0) u_n(r) dr.
 \label{3c}
\end{equation}
and the functions $M(y_n)$, the so named Moshinsky functions, are defined as \cite{gc10}
\begin{equation}
M(y_n)=\frac{i}{2\pi}\int_{-\infty}^{\infty}\frac{{\rm e}^{ik(r-a)}{\rm e}^{-ik^2t}}{k-\kappa_n}dk=
\frac{1}{2}{\rm e}^{i(r-a)^2/4 t} w(iy_n),
\label{16c}
\end{equation}
where
\begin{equation}
y_n={\rm e}^{-i\pi /4}(1/4t)^{1/2}[(r-a)-2 \kappa_nt],
\label{17c}
\end{equation}
and the function $w(z)=\exp(-z^2)\rm{erfc(-iz)}$ in (\ref{16c}) stands for the Faddeyeva or complex error function \cite{abramowitz} for which there exist efficient computational tools to calculate it \cite{poppe}. The argument $y_n^{\circ}$ of the functions $M(y_n^0)$ in (\ref{b6}) is that of $y_n$ given by (\ref{17c}) with $r=a$, namely,
\begin{equation}
y^\circ_n=-{\rm e}^{-i\pi /4}\kappa_nt^{1/2}.
\label{17d}
\end{equation}

Assuming that the initial state $\Psi(r,0)$ is normalized to unity, it follows from the closure relation (\ref{9y}) that,
\begin{equation}
{\rm Re}\sum_{n=1}^\infty \left\{ C_n \bar{C}_n\right\}= 1,
\label{9z}
\end{equation}
where
\begin{equation}
{\bar C}_n=\int_0^a \Psi^*(r,0) u_n(r) dr.
\label{3e}
\end{equation}
Equation (\ref{9z}) indicates that ${\rm Re}\,\{C_n{\bar C}_n\}$ cannot be interpreted as a probability, since in general it is not a positive definite quantity. Nevertheless, one may see that it represents the `strength'  or `weight' of the initial state in the corresponding resonant state. One might see the coefficients  ${\rm Re}\,\{C_n{\bar C}_n\}$ as some sort of quasi-probabilities.

The solution $\Psi(r,t)$ for $r \leq a$,  given by the first equation in (\ref{b6}), is the relevant ingredient to calculate the survival and nonescape probabilities, as discussed in \cite{gcmv12,gcmv07,gc10}. For $r \geq a$, the solution $\Psi(r,t)$, given by the second equation in (\ref{b6}),  describes the propagation of a single decaying particle along the external region. This has been discussed in \cite{gcmv12,gcmv13,gc10}.

\section{Decay of two identical noninteracting particles}

For identical  non interacting particles, it is well known that the Hamiltonian $H$ must be symmetric under the permutation of the indices of the particles so the exchange operator and $H$ necessarily commute. It is enough to impose the appropriate symmetry or antisymmetry conditions on the initial state $\Psi(y_1,y_2,0)$ since symmetry is conserved as time evolves. As a consequence, the time evolution for decay of two identical particles may be written as,
\begin{equation}
\Psi(r_1,r_2,t)= \int_0^a {\!\int_0^a {\!g(r_1,y_1,t)g(r_2,y_2,t)\Psi(y_1,y_2,0)\,dy_1}\,dy_2}.
\label{d1}
\end{equation}

A simple choice corresponding to a symmetric state, is given by the product
of single particle states $\psi_\alpha(y_1,0)$ and $\psi_\alpha(y_2,0)$, with $\alpha$ denoting the corresponding initial state,
\begin{equation}
\Psi_{\alpha,\alpha}(y_1,y_2,0)=\psi_\alpha(y_1,0)\psi_\alpha(y_2,0).
\label{d2}
\end{equation}
Substitution of (\ref{d2})  into (\ref{d1}) allows us to write the decaying \textit{factorized symmetric} state
as a product  of two evolving single particle decaying states,
\begin{equation}
\Psi_{\alpha,\alpha}(r_1,r_2,t)=\psi_\alpha(r_1,t)\psi_\alpha(r_2,t).
\label{d3}
\end{equation}
Using (\ref{b6}) one may write the resonance expansion for $\Psi(r_1,r_2,t)$ as,

\begin{equation}
\Psi_{\alpha,\alpha}(r_1,r_2,t)=\sum_{p,q=-\infty}^{\infty}
\left \{ \begin{array}{cc}
\left [C_{p,\alpha}u_p(r_1)M(y^\circ_p)\right ] \left [C_{q,\alpha}u_q(r_2)M(y^\circ_q)\right ], & \quad  (r_1,r_2) \leq a \\[.4cm]
\left [C_{p,\alpha}u_p(a)M(y_p)\right ]\left [C_{q,\alpha}u_q(a)M(y_q)\right ],
& \quad (r_1,r_2) \geq a,
\end{array}
\label{d4}
\right.
\end{equation}
where  $C_{n,\alpha}$, with $n=p$ or $q$, is given by,
\begin{equation}
C_{n,\alpha}=\int_0^a {\!u_n(y)\psi_\alpha(y)\,\mathrm{d}y}.
\label{d5}
\end{equation}
For different initial states for each particle, it follows from the Pauli Principle that initial two-particle state must consist of a linear combination of single-particle states corresponding to the symmetric and antisymmetric combinations. Denoting by  $\psi_s(y_1,0)$ and $\psi_s(y_2,0)$ the initial states, with $s=\alpha$ or $\beta$, one may write,
\begin{equation}
\Psi_{\alpha,\beta}(y_1,y_2,0)=\frac{1}{\sqrt{2}}\left [\psi_\alpha(y_1,0)\psi_\beta(y_2,0)\pm \psi_\beta(y_1,0)\psi_\alpha(y_2,0)\right],
\label{d6}
\end{equation}
where respectively, the plus sign refers to \textit{entangled symmetric} and the minus sign to
\textit{entangled antisymmetric} states. It follows immediately that the corresponding time evolved state reads,
\begin{equation}
\Psi_{\alpha,\beta}(r_1,r_2,t)=\frac{1}{\sqrt{2}}\left [\psi_\alpha(r_1,t)\psi_\beta(r_2,t)\pm \psi_\beta(r_1,t)\psi_\alpha(r_2,t)\right ].
\label{d7}
\end{equation}
Then, using (\ref{b6}) one may write the two-particle decaying wave function as
\begin{equation}
\Psi_{\alpha,\beta}(r_1,r_2,t)=\sum_{p,q=-\infty}^{\infty}
\left \{ \begin{array}{cc}
\left [C_{p,\alpha}C_{q,\beta} \pm C_{p,\beta}C_{q,\alpha}\right ] u_p(r_1)u_q(r_2)M(y^\circ_p)M(y^\circ_q),
& \,\,  (r_1,r_2) \leq a \\[.4cm]
\left [C_{p,\alpha}C_{q,\beta} \pm C_{p,\beta}C_{q,\alpha}\right ] u_p(r_1)u_q(a)M(y^\circ_p)M(y_q),
& \,\,  r_1 \leq a,\, r_2 \geq a \\[.4cm]
\left [C_{p,\alpha}C_{q,\beta} \pm C_{p,\beta}C_{q,\alpha}\right ] u_p(a)u_q(a)M(y_p)M(y_q),
& \,\, (r_1,r_2) \geq a,
\end{array}
\label{d8}
\right.
\end{equation}
where the coefficients $C_{n,s}$ with $n=(p,q)$ or $s=(\alpha,\beta)$ may be figured out from  (\ref{d5}).
It is worth recalling that the coefficients $\{C_{n,s}\}$, which involve only single-particle states,
fulfill the relationship \cite{gc10}
\begin{equation}
{\rm Re} \left( \sum_{n=1}^\infty C_{n,s}{\bar C}_{n,s} \right )=1.
\label{sumrule}
\end{equation}
In the above expression  ${\bar C}_{n,s}$ is defined as (\ref{d5}) with $\psi_s(y)$ substituted by $\psi^*_s(y)$. Hence for real initial states,  ${\bar C}_{n,s}=C_{n,s}$.

The two-particle solution given by the first equation in (\ref{d8}) permits to calculate the corresponding two-particle expressions for the survival and nonescape probabilities \cite{gcm11,gcr17}.
The situation described by the second equation in (\ref{d8}), which may be used to obtain the probability density where one particle remains within the interaction region whereas the other propagates along the external region of the interaction is, since both particles decay, presumably a very small quantity. In any case, here we shall be concerned with the situation provided by the last equation in (\ref{d8}) to obtain the probability density where both decaying particles propagate along the external region of the interaction.

\section{Models}

In this section we consider the s-wave delta-shell potential and the double barrier resonant potential to illustrate the profile of the probability density for two identical non interacting decaying particles as a function of time and fixed values of the corresponding positions. In particular, we are interested in how the resonance spectra affects the profile of the probability density.

\subsection{Delta-shell potential}

The delta-shell potential is defined as,
\begin{equation}
V(r)=\lambda\delta(r-a),
\label{eq}
\end{equation}
where $\lambda$ stands for the intensity of the potential and $a$ for the radius. We model the initially confined state by a given infinite box state,
\begin{equation}
\Psi(r,0)=\sqrt{\frac{2}{a}} \, \sin \left (\frac{q\pi}{a}r \right ), \quad q=1,2,3...
\label{is}
\end{equation}
The resonance solutions to the problem with complex energy eigenvalues $\kappa_n^2=\mathcal{E}_n-i\Gamma_n/2$ read,
\begin{equation}
u_n(r)=\left\{
\begin{array}{cc}
A_n \,\sin (\kappa_nr) & r\leq a \\[.4cm]
B_n \,e^{i\kappa_nr}, & r \geq a,
\end{array}
\right.
\label{5c}
\end{equation}
where $\kappa_n=\upsilon_n-i\delta_n$.
From the continuity of the above solutions and the discontinuity of its derivatives with respect to $r$ (due to the $\delta$-function interaction) at the boundary value $r=a$, it is obtained that the set of $\kappa_n$'s satisfy the equation,
\begin{equation}
2i\kappa_n + \lambda ( e^{2i\kappa_na}-1)=0.
\label{5d}
\end{equation}
For $\lambda > 1$ one may write the approximate analytical solutions to Eq. (\ref{5d}) as \cite{gc10}
\begin{equation}
\kappa_n \approx \frac{n\pi}{a} \left (1-\frac{1}{\lambda a}\right ) -i \,\frac{1}{a}\left( \frac{n\pi}{\lambda a}\right )^2.
\label{5e}
\end{equation}
Using iterative procedures as the Newton-Rapshon method, one may obtain the poles $\kappa_n$ with the desired degree of approximation.
It is worth noticing that as the intensity of the potential $\lambda \rightarrow \infty$, the complex poles $\kappa_n=\upsilon_n-i\delta_n$ tend to the real infinite box eigenfunctions, and similarly, the resonance eigenfunctions $u_n(r)$ tend to the infinite box model eigenfunctions. This implies that for a large finite value of the intensity $\lambda$, an  initial infinite box state $\Psi(r,0)$ with $q=m$, has a larger overlap with the resonant state $u_m(r)$ than with any other resonant state. For a given value of the intensity $\lambda$ and a radius $a$ of the $\delta$-potential, one may then  evaluate the corresponding set of complex poles $\{\kappa_n\}$, the set of normalized resonance states $\{u_n(r)\}$,  and the  expansion coefficients $\{C_n\}$ to the problem.

\begin{figure}[!tbp]
\centering
\begin{minipage}{0.45\textwidth}
\centering
\includegraphics[width=1.1\textwidth]{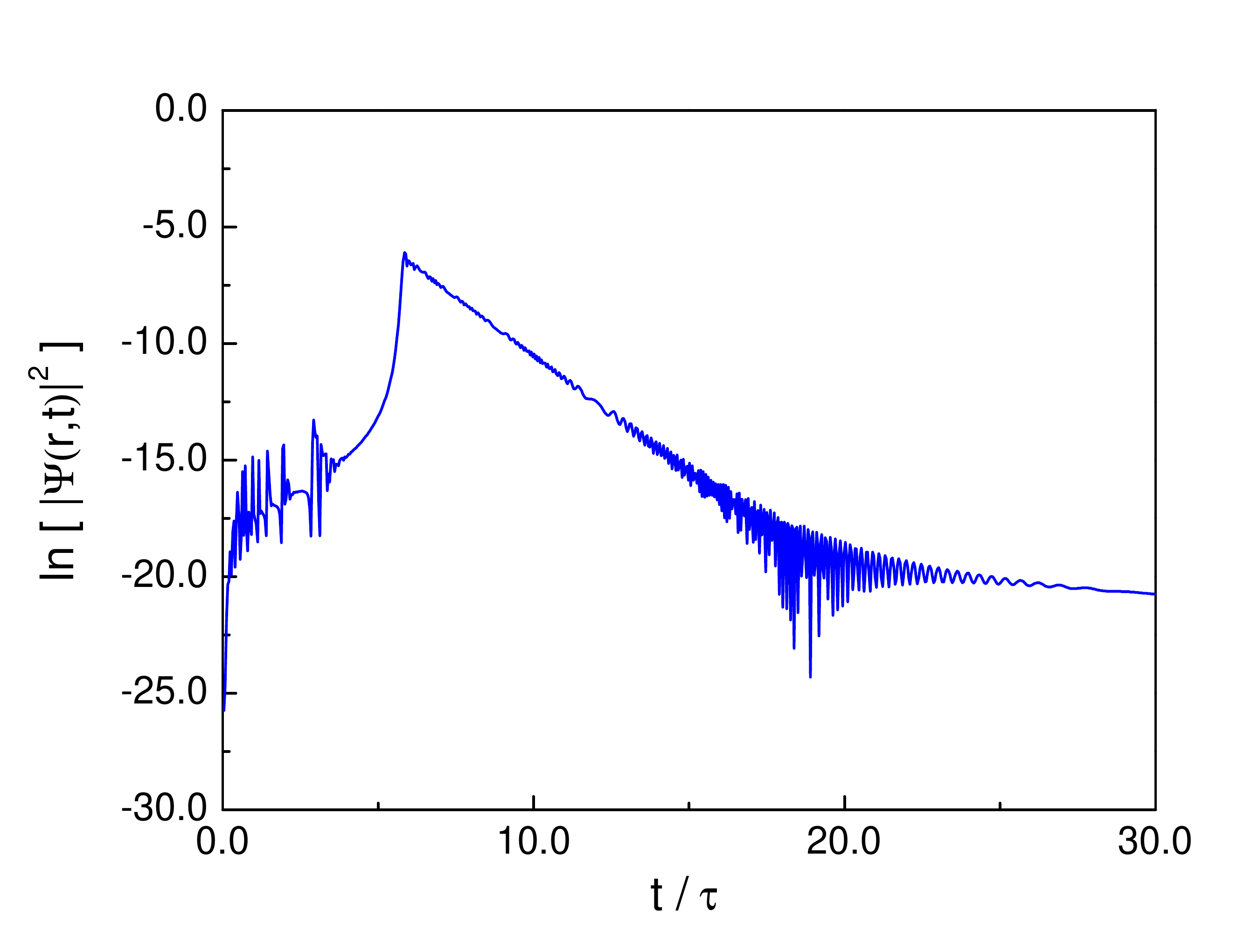}
\caption{$\ln [|\Psi(r,t)|^2]$ vs time in lifetime units for single-particle decay. See text.}
\label{figure1}
\end{minipage}\hfill
\begin{minipage}{0.45\textwidth}
\centering
\includegraphics[width=1.1\textwidth]{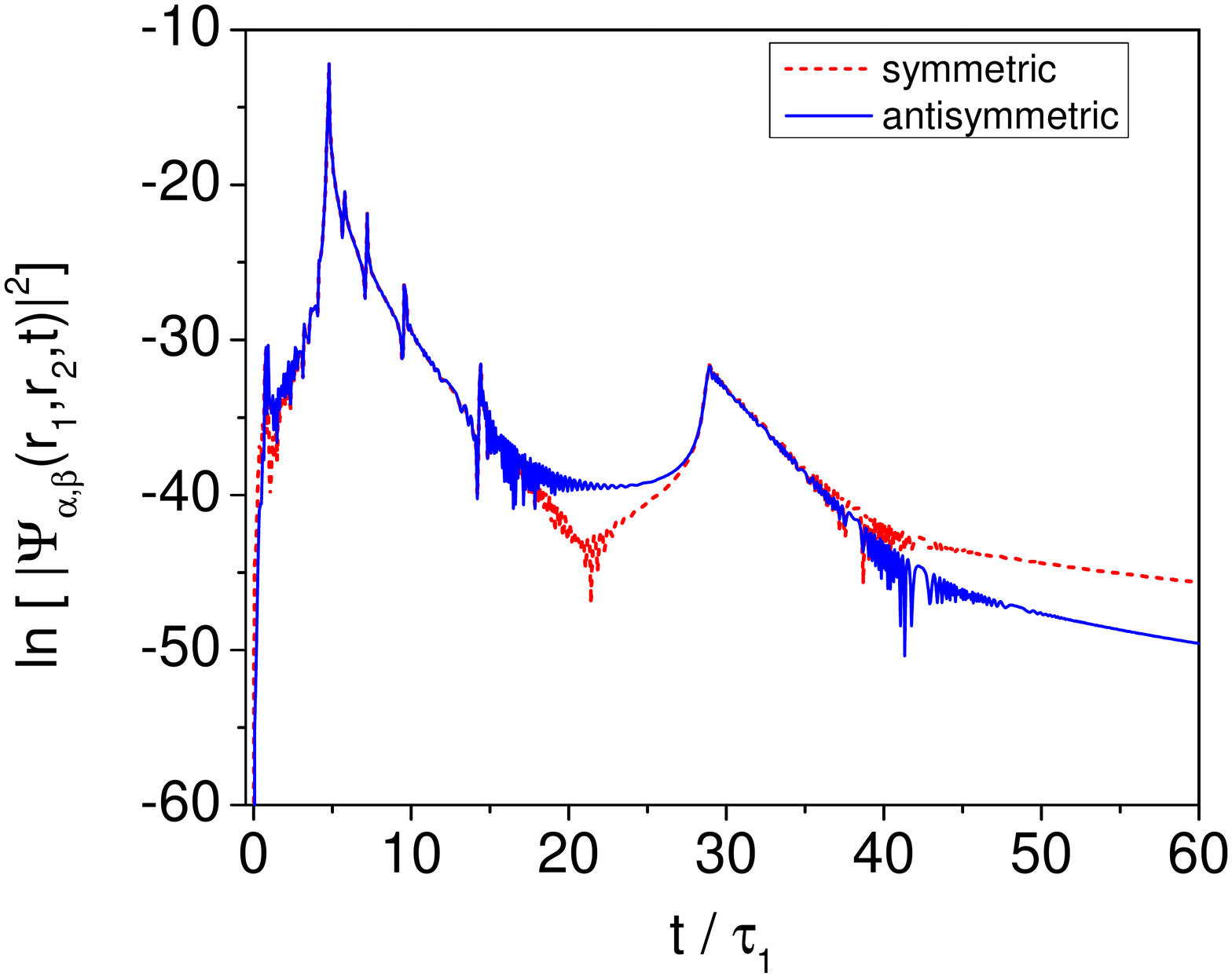}
\caption{$\ln [|\Psi_{\alpha,\beta}(r,t)|^2]$ vs time in lifetime units for two-particle decay of symmetric and antisymmetric states. See text.}
\label{figure2}
\end{minipage}
\end{figure}

Figure \ref{figure1} exhibits a plot of  $\ln [|\Psi(r,t)|^2]$, given by the second equation of equation of (\ref{b6}),  as a function of time in lifetime  units, calculated at the distance $r=3000\,a$. Here we choose $q=1$, and hence the real part of the overlap of the initial state with the resonant state  with  $n=1$, in view of (\ref{9z}), is of the order of unity. We choose the potential parameters  $\lambda=100$ and $a=1$. As time evolves we observe a buildup of the probability density characterized by resonance peaks that are due to high resonance contributions until it reaches the highest peak which corresponds to the dominating contribution that arises from the resonance term with $n=1$ in  the second equation of  (\ref{b6}). This is corroborated by the fact that the highest peak is given by $t_1 \approx r/2\upsilon_1$, which follows by inspection of the argument (\ref{17c}) of the Moshinsky function. As time evolves further, the probability density mimics the behavior of the survival or nonescape probabilities. There is a regime of exponential decay  which is followed by the transition to nonexponential decay where the probability density  evolves as the inverse power of time $t^{-3}$ \cite{gcmv13}.
\begin{figure}[!tpb]
\centering
\begin{minipage}{0.45\textwidth}
\centering
\includegraphics[width=1.1\textwidth]{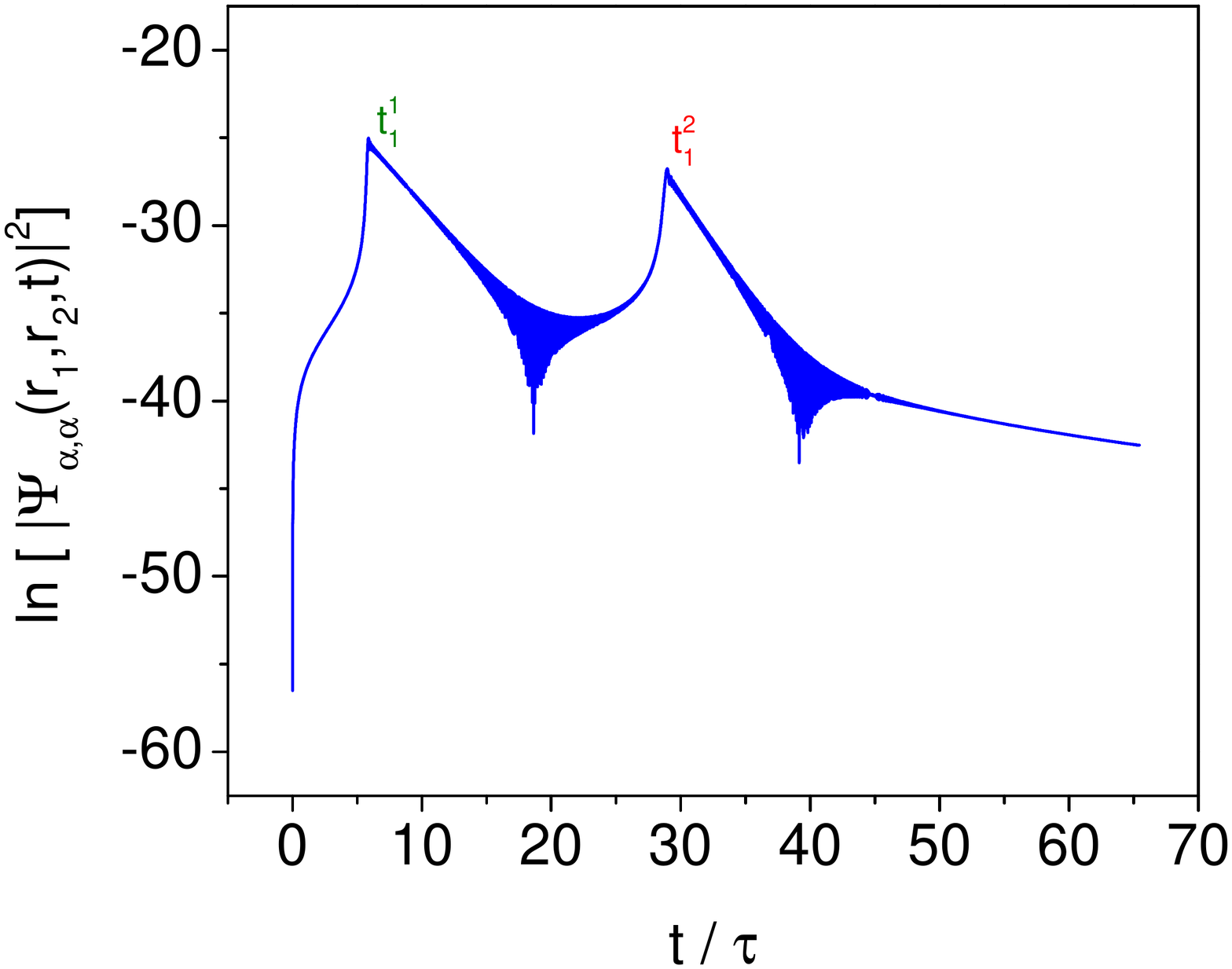}
\caption{$\ln [|\Psi_{\alpha,\alpha}(r,t)|^2]$ vs time in lifetime units for single-particle decay. See text.}
\label{figure3}
\end{minipage}\hfill
\begin{minipage}{0.45\textwidth}
\centering
\includegraphics[width=1.1\textwidth]{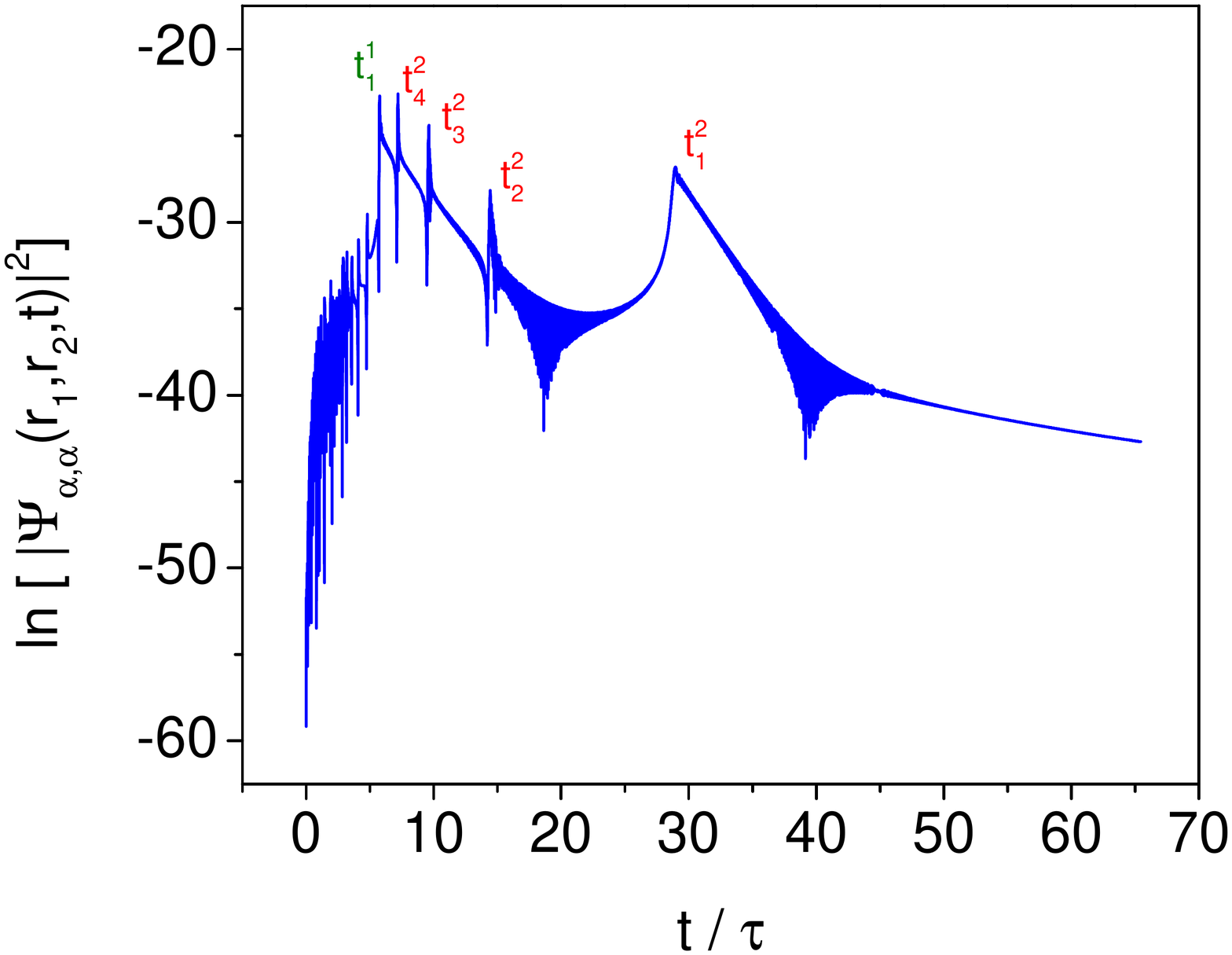}
\caption{$\ln [|\Psi_{\alpha,\alpha}(r,t)|^2]$ vs time in lifetime units for two-particle decay. See text.}
\label{figure4}
\end{minipage}
\end{figure}

Figure \ref{figure2} displays the plot of $\ln [|\Psi_{\alpha,\beta}(r,t)|^2]$ as a function of time in lifetime units for the two-particle probability density  given by the third equation of (\ref{d8}) with   $r_1=2400\,a$  and $r_2=15000\,a$, for the delta-shell potential with parameters $\lambda=100$ and $a=1$ with $\alpha=1$,  and $\beta=6$, which correspond, respectively to $q=1$ and $q=6$ of equation (\ref{is}). One sees that the probability density profile  is very similar for symmetric and antisymmetric states. This seems to be a common feature of the propagating solutions. This is not so along the internal interaction region \cite{gcm11}. Essentially the symmetric and antisymmetric quantities differ in the valley region between the two broad peaks and in the corresponding asymptotic behavior at long times. It turns out that the symmetric state goes as  $t^{-6}$ and the antisymmetric one as $t^{-10}$, as obtained along the internal region \cite{gcm11}. Comparison of figures \ref{figure1} and \ref{figure2} suggests that the two broad resonance peaks are related to the decay of the particles in the initial states $\alpha=1$ and $\beta=6$. This is corroborated by calculating the corresponding  peak valu in time domain for the state $\alpha=1$, $t_1^1=r_1/2\upsilon_1$, and for the state $\beta=6$, given by  $t_6^2=r_2/2 \upsilon_6$. Further inspection of figure \ref{figure2} suggests that the resonance peaks in time domain
that seem to disrupt the decaying part of the first broad peak correspond to high energy resonances of the system.

\begin{figure}[!tpb]
\centering
\begin{minipage}{0.45\textwidth}
\centering
\includegraphics[width=1.1\textwidth]{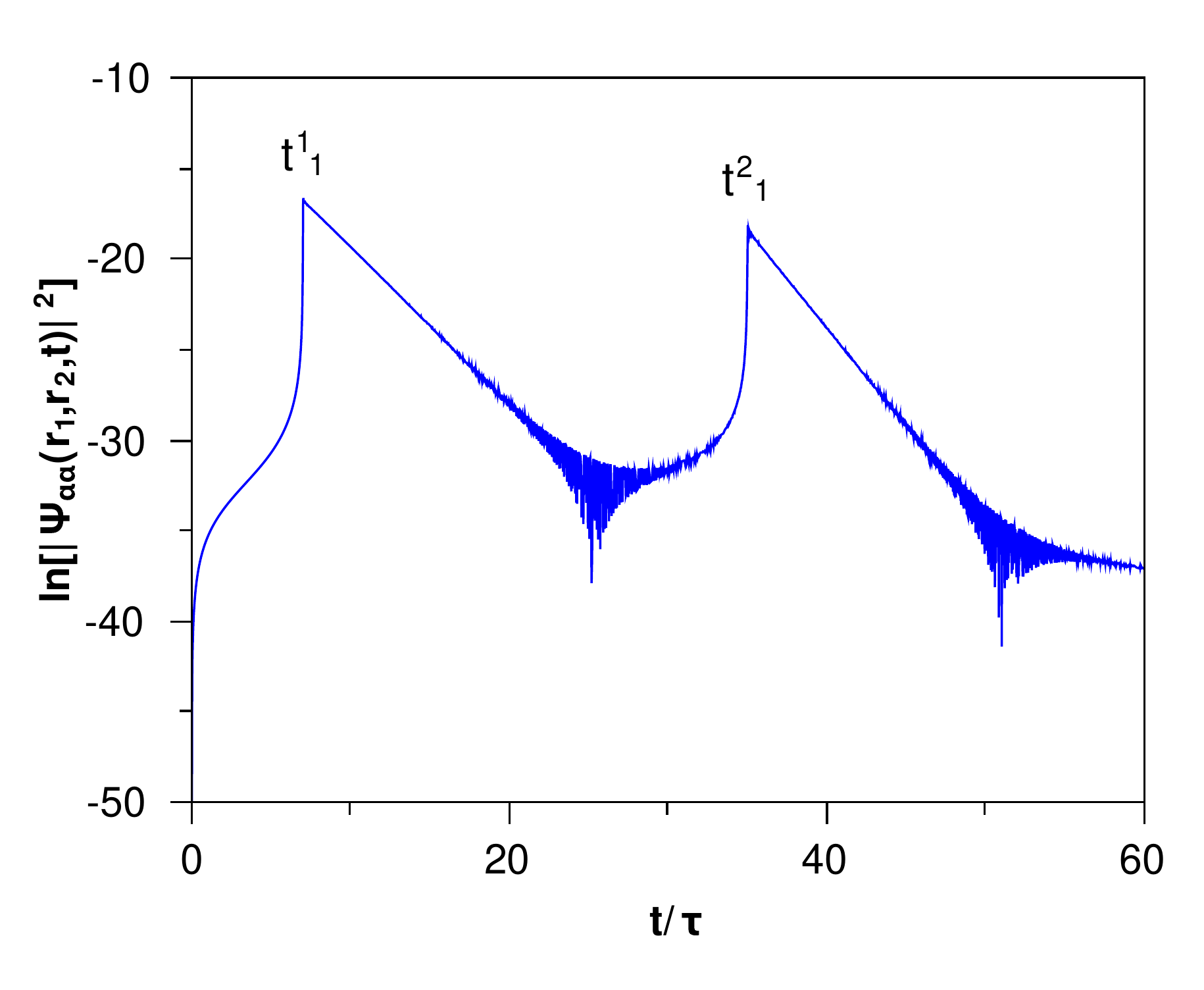}
\caption{$\ln [|\Psi_{\alpha,\alpha}(r,t)|^2]$ vs time in lifetime units for single-particle decay. See text.}
\label{figure5}
\end{minipage}\hfill
\begin{minipage}{0.45\textwidth}
\centering
\includegraphics[width=1.1\textwidth]{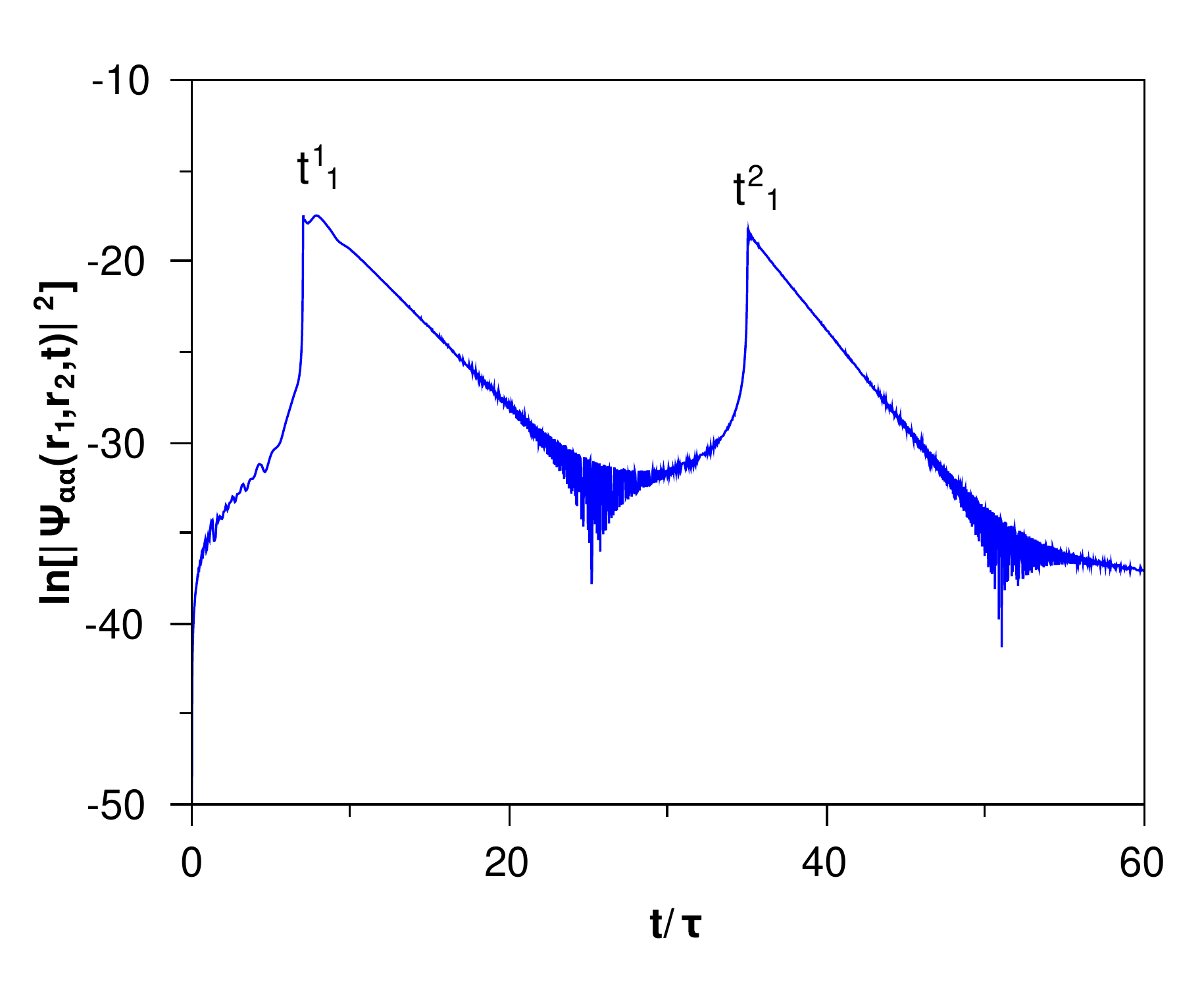}
\caption{$\ln [|\Psi_{\alpha,\alpha}(r,t)|^2]$ vs time in lifetime units for two-particle decay. See text.}
\label{figure6}
\end{minipage}
\end{figure}
This is further corroborated by figures \ref{figure3} and \ref{figure4}, which display the plot of the
$\ln [|\Psi_{\alpha,\alpha}(r,t)|^2]$ as a function of time in lifetime units for a symmetric state with $\alpha=1$
with $r_1=3000\,a$ and $r_2=5r_1$. In figure \ref{figure3} both states correspond to  $\alpha_1$, namely, (\ref{is}) with $q=1$.  One sees two broad peaks, one peaked at $t_1^1 =r_1/2\upsilon_1$ and the other at $t_1^2 = r_2/2\upsilon_1$. In fact in lifetime units  $t_1^1=5.73$ with  and $t_1^2=28.68$ with $\upsilon_1=3.11052$ The lifetime is $\tau=82.058$. The probability density is larger around these peaked values. On the other hand, figure \ref{figure4} refers to the same system with the contribution of $1000$ resonance poles. One sees that a number of peaks that arise from higher resonance terms at $t_2^2$, $t_3^2$ and $t_4^2$, disrupt the exponential decaying regime corresponding to the first broad peak. In lifetime units these times are given by: $t_2^2=14.34$, $t_3^2=9.56$ and $t_4^2=7.17$. The corresponding $\upsilon$'s read:$\upsilon_2=6.2213$, $\upsilon_3=9.3325$,$\upsilon_4=12.4444$.

\subsection{Double barrier system}
The delta potential in general corresponds to a situation that involves many sharp isolated resonances poles, namely, $\upsilon_n \gg \delta_n$. It is good model in systems with these characteristics. However, there are systems with just a few sharp resonances, may be two, which refer to particles trapped  by barrier of finite height, where the
high energy resonances are not trapped and hence are broad and presumably overlapping. A model with these characteristics is a double barrier with a well in between. The calculation  of the complex poles have been discussed elsewhere \cite{gcv06} and there is no space to present it here. It follows a procedure analogous to the previous model. It should be noted that since the potential is one dimensional, the normalization has an additional term.

As in the previous two figures, our aim is to make a comparison  with a situation where the double barrier
system holds two sharp resonances and the remaining, the higher resonance energies, are situated above the barrier height. The system parameters are: barrier heights $V=40$, well width $w=1$, and barrier widths $b=1$. Figures \ref{figure5} and \ref{figure6} exhibit each a plot of $\ln [|\Psi_{\alpha,\alpha}(r,t)|^2]$ as a function of time in lifetime units with $r_1=600 000$ and $r_2=5r_1$. The Since there are only two resonance states, it is not surprising to obtain a profile for the decaying probability density similar to that in figure \ref{figure3}. However, in contrast to figure \ref{figure4}, the many resonance calculation, up to $50$ states presented in figure \ref{figure6} remains almost identical to that involving just the two sharp resonances situated below the barrier heights. In this case, in lifetime units $t_1^1 = 7.0$ and $t_2^1=35.0$. The corresponding pole is $\upsilon_1=2.3725$ and the lifetime $\tau=18 067.23$.
The above result seems to imply that broad overlapping resonances do not play a relevant role in the profile of the decaying probability density for two particles.

\section{Concluding remarks}

We find of interest the result that a multilevel sharp resonance spectra may disrupt the exponential decaying regime which occurs in the case of a multilevel broad resonance spectra. Recent experiments  involving ultracold gases \cite{serwane11} may be used to test this novel nonexponential behavior in the time evolution of decay of identical noninteracting particles. Another result worth noticing is that symmetric and antisymmetric states evolve almost identically except in the small valley region between the two broad peaks related to the initial resonance states and in the long time asymptotic regime.

\section*{Acknowledgments}
G.G-C acknowledges the partial financial support of UNAM-DGAPA-PAPIIT IN105618. R.R.  acknowledges financial support from PRODEP under the program {\it Apoyo para Estancias Cortas de Investigación}, Convocatoria 2018. R.R. also thanks Instituto de Física de la UNAM for its hospitality.

\section*{References}

\begin{thebibliography}{35}
%
\bibitem{gcmv12} Garc\'{\i}a-Calder\'on G, M\'attar A and Villavicencio J 2012 \emph{Phys. Scr.} {\bf T151} 1

\bibitem{gcmv07} Garc\'{\i}a-Calder\'on G, Maldonado I and Villavicencio J 2007 \emph{Phys. Rev. A} {\bf 76} 012103

\bibitem{gcmv13} Garc\'{\i}a-Calder\'on G,  Maldonado I and Villavicencio J 2013 \emph{Phys Rev A} {\bf 88} 052114

\bibitem{wigner30} Weisskopf V F and Wigner E P 1930  {\it Z. Phys.} {\bf 65} 54

\bibitem{sakaki87} Tsuchiya M, T. Matsusue T and Sakaki H 1987 {\it Phys. Rev. Lett.} \textbf{59} 2356

\bibitem{serwane11}  Serwane F, Z\"urn G , Lompe T, Ottenstein T B, Wenz  A N and Jochim S 2011
{\it Science } \textbf{332} 336

\bibitem{gcv06} Garc\'{\i}a-Calder\'on G and Villavicencio J 2006 \emph{Phys. Rev A} {\bf 73} 062115

\bibitem{gcr16} Garc\'{\i}a-Calder\'on G and Romo R 2016 \emph{Phys Rev. A} {\bf 93} 022118

\bibitem{gcm11}  Garc\'{\i}a-Calder\'on G and Mendoza-Luna L G 2011 \emph{Phys. Rev. A} {\bf 84} 032106

\bibitem{gc10} Garc\'{\i}a-Calder\'on G 2010 \emph{Adv. in Quant. Chem.} {\bf 60} 407

\bibitem{gcp76}  Garc\'{\i}a-Calder\'on G and  Peierls R 1976 \emph{Nucl.Phys. A} {\bf 265} 443

\bibitem{newton} Newton R G 2002 {\it Scattering Theory of Waves and Particles} (New York: Dover Publications) chapter 12

\bibitem{abramowitz}Abramowitz M and Stegun A I 1964 \emph{ Handbook of Mathematical
Functions} (New York: Dover Publications)

\bibitem{poppe} Poppe G M P  and  Wijers C M J 1990 \emph{ACM Trans. Math. Softw.} {\bf 16} 38

\bibitem{gc15} Garc\'{\i}a-Calder\'on G 2015 \emph{J. Phys. Conf. Series} {\bf 626} 012064 (arXiv:quant-ph/1812.09319)

\bibitem{gcr17} Garc\'{\i}a-Calder\'on G and Romo R 2017 \emph{Phys. Rev. A} {\bf 96} 062124

%
\end{thebibliography}
\end{document}